\documentclass[12pt]{emulateapj}

\usepackage[dvips]{color}

\shorttitle{M87 jet base}
\shortauthors{Kino et al.}

\begin{document}

\title{Relativistic electrons and magnetic field of the M87 jet
on $\sim$ten Schwarzschild radii scale}
\author{
M. Kino\altaffilmark{1,2},
F. Takahara\altaffilmark{3},
K. Hada\altaffilmark{4,5},
A. Doi\altaffilmark{1}}

\altaffiltext{1}{
Institute of Space and Astronautical Science, Japan Aerospace Exploration Agency, 
3-1-1 Yoshinodai, 229-8510 Sagamihara, Japan} 
\email{kino@vsop.isas.jaxa.jp}
\altaffiltext{2}{Korea Astronomy and Space Science Institute,
776 Daedukdae-ro, Yusong, Daejon 305-348, Korea}
\altaffiltext{3}{Department of Earth and Space Science,
Osaka University, Toyonaka 560-0043, Japan}
\altaffiltext{4}{
INAF - Istituto di Radioastronomia, 
via Gobetti 101, 40129 Bologna, Italy}
\altaffiltext{5}{National Astronomical Observatory of Japan
                 2-21-1 Osawa, Mitaka, Tokyo, 181-8588, Japan}



\begin{abstract}

We explore energy densities of
magnetic field and relativistic electrons in 
the M87 jet.
Since the radio core at the jet base is 
identical to the optically thick surface against
synchrotron self absorption (SSA),
the observing frequency is identical to
the SSA turnover frequency. 
As a first step, we assume the radio core 
as a simple uniform sphere geometry.
Using the observed angular size of the radio core
measured by the Very Long Baseline Array at 43~GHz, 
we estimate the energy densities of magnetic field ($U_{B}$)
and relativistic electrons ($U_{e}$) 
based on the standard SSA formula.
Imposing the condition that the Poynting power
and relativistic electron one should be smaller than
the total power of the jet,
we find that 
(i) the allowed range of 
the magnetic field strength ($B_{\rm tot}$) is  
$1~{\rm G} \le B_{\rm tot} \le 15~{\rm G}$, 
and that
(ii)  
$ 1\times 10^{-5} \le U_{e}/U_{B} \le 6 \times 10^{2}$ 
holds.
The uncertainty of $U_{e}/U_{B}$ comes from  
the strong dependence  on the angular size of the radio core
and the  minimum Lorentz factor of non-thermal electrons ($\gamma_{e,\rm min}$)
in the core.
It is still open that the
resultant energetics is consistent with either
the magnetohydrodynamic jet
or with kinetic power dominated jet even on
$\sim 10$ Schwarzschild radii scale.

\end{abstract}

\keywords{galaxies: active --- galaxies: jets --- 
radio continuum: galaxies}

\section{Introduction}
\label{sec:intro}

Formation mechanism  of relativistic jets
in active galactic nuclei (AGNs) remains 
as a longstanding unresolved problem in astrophysics.
Although the importance of 
magnetic field energy density ($U_{\rm B}$)
and 
relativistic electron one ($U_{e}$)  
for resolving the formation mechanism has been 
emphasized  
(e.g., Blandford and Rees 1978),
it is not observationally clear 
whether either $U_{\rm B}$ or $U_{e}$
is dominant at the jet base.
Relativistic magnetohydrodynamics
models for relativistic jets generally
assume highly magnetized plasma at the jet base
(e.g., 
Koide et al. 2002;
Vlahakis and Konigl 2003; 
McKinney and Gammie 2004;
Krolik et al. 2005;
McKinney 2006; 
Komissarov et al. 2007;
Tchekhovskoy et al. 2011;
Toma and Takahara 2013;
Nakamura and Asada 2013),
while an alternative model 
assumes a pair plasma dominated 
``fireball"-like state at the jet base
(e.g., 
Iwamoto and Takahara 2002;
Asano and Takahara 2009 and reference therein).
Although deviation from  equi-partition 
(i.e., $U_{e}/U_{B}\approx 1$)
is essential for investigation of 
relativistic jet formation,
none has succeeded in obtaining  a robust estimation of
$U_{e}/U_{B}$ at the jet base.

M87, a nearby giant radio galaxy located at 
a distance of $D_{\rm L}=16.7~{\rm Mpc}$ (Jordan et al. 2005), 
hosts one of the 
most massive super massive black hole
$M_{\bullet}=(3-6)\times 10^{9}~M_{\odot}$ 
(e.g., 
Macchetto et al. 1997; 
Gebhardt and Thomas 2009; 
Walsh et al. 2013).
Because of the largeness of 
the angular size of its central black hole,
M87 is well known as
the best source for imaging the deepest part of 
the jet base (e.g., Junor et al. 1999).
Furthermore, M87 has been well studied
at wavelengths from radio to 
Very High Energy (VHE) $\gamma$-ray
(Abramowski et al. 2012; Hada et al. 2012 
and reference therein)
and causality arguments based on 
VHE $\gamma$-ray outburst in February 2008
indicate that the VHE emission region
is less than $\sim 5\delta~{\rm R_{s}}$ where 
$\delta$ is the relativistic Doppler factor (Acciari et al. 2009).
The Very-Long-Baseline-Array (VLBA)
beam resolution at 43~GHz typically attains
about $0.21\times 0.43~{\rm mas}$
which is equivalent to
$5.3\times 10^{16}
\times1.1\times 10^{17} ~{\rm cm}$.
When $M_{\bullet}=6\times 10^{9}~M_{\odot}$ holds
(Gebhardt et al. 2009),
then VLBA beam resolution approximately corresponds 
to $30\times 60~{\rm R_{s}}$.
Recent progresses of 
Very-Long-Baseline-Interferometry (VLBI) observations
have revealed the inner jet structure, 
i.e., 
frequency and core-size relation, and
distance and core-size relation down to
close to 
$\sim 10$ Schwarzschild radii ($R_{\rm s}$) scale
(Hada et al. 2011, hereafter H11).
Thus, the jet base of M87 is the best laboratory
for investigations of $U_{e}/U_{B}$ in
the real
vicinity of the central engine.

Two significant forward steps are recently
obtained in M87 observations which motivate the present work.
First,
Hada et al. (2011) succeeded in directly measuring 
core-shift phenomenon at the jet base of M87
at 2, 5, 8, 15, 
24 and 43 GHz.
The radio core position at each frequency has
been obtained by the astrometric observation (H11).
Since the radio core surface corresponds 
to the optically-thick surface at each frequency,
the synchrotron-self-absorption (SSA) 
turnover frequency $\nu_{\rm ssa}$ is identical
to the observing frequency itself.
\footnote{
Difficulties
for applying the basic SSA model 
to real sources has been already recognized
 by several authors
(
Kellermann and Pauliny-Toth 1969;
Burbidge et al. 1974; 
Jones et al. 1974a, 1974b;
Blandford and Rees 1978;
Marscher 1987) 
due to insufficiently accurate determination 
of $\nu_{\rm ssa}$ and $\theta_{\rm obs}$.}
Second,
we recently measure core sizes
in Hada et al. (2013a) (hereafter H13). 
Hereafter we focus on the radio core at 43~GHz.
In H13, we select VLBA data observed after 2009 
with  sufficiently good qualities 
(all 10 stations participated and good uv-coverages).
To measure the width of the core,
a single, full-width-half-maximum (FWHM) 
Gaussian is fitted for the observed radio 
core at 43~GHz
in the perpendicular direction to the jet axis
and we derive the width of the core ($\theta_{\rm FWHM}$).
We stress that the core width is free from the 
uncertainty of viewing angle.
Therefore, using $\theta_{\rm FWHM}$ at 43~GHz,
we can estimate values of
$U_{e}/U_{B}$ in the 43~GHz core of M87
for the first time.


In section 2, we derive an explicit form
of $U_{e}/U_{B}$ by using the standard 
formulae of synchrotron absorption processes.
As a first step,
we simplify a geometry of the radio core
as a single uniform sphere although the real 
geometry is probably more complicated.
In section 3, 
we 
estimate $U_{e}/U_{\rm B}$ in
the M87 jet base
 by using the VLBA data at 43~GHz obtained in H13.
In section 4, we summarize the result and 
discuss relevant implications.
In this work, we define the radio spectral index $\alpha$
as $S_{\nu}\propto \nu^{-\alpha}$
and 
we assume $M_{\bullet}=6\times 10^{9}~M_{\odot}$.

\section{Model}

Here, we derive explicit expressions of 
the strength of total magnetic field $B_{\rm tot}$ and
$U_{e}/U_{B}$.
Several papers have extensively 
discussed the determination of 
magnetic field strength $B_{\rm tot}$.
Fundamental formulae of SSA processes are shown
in the following references and we follow them 
(Ginzburg and Syrovatskii 1965, hereafter GS65; 
Blumenthal and Gould 1970, hereafter BG70;
Pacholczyk 1970,  
Rybicki and Lightman 1979, hereafter RL79).
Here, we will show a simple
derivation of the explicit expressions of 
$B_{\rm tot}$ and $U_{e}/U_{B}$ with sufficient accuracy.



\subsection{Method}

For clarity, we briefly
summarize the method for 
determining  $B_{\rm tot}$ and $U_{e}/U_{B}$ in advance.
The theoretical unknowns related to the magnetic field and 
relativistic electrons in the observed radio core with 
its angular diameter $\theta_{\rm obs}$ are following four;   
$B_{\rm tot}$, 
$K_{e}$ (the normalization factor of non-thermal electron number density), 
$\gamma_{e,\rm min}$ (the minimum Lorentz factor of non-thermal electrons)
\footnote{The maximum Lorentz factor of non-thermal electrons ($\gamma_{e,\rm max}$)
is not used since the case of $p>2$ is considered in this work
based on the ALMA observation (Doi et al. 2013).}, and
$p$ (the spectral index of non-thermal electrons).
Among them,  
$\gamma_{e,\rm min}$, and $p$
are directly constrained by radio observations at mm/sub-mm wavebands.
The remaining $B_{\rm tot}$ and $K_{e}$ can be solved by 
using the two general relations  which hold at $\nu=\nu_{\rm ssa}$ 
shown in Eqs.~(\ref{eq:tau}) and (\ref{eq:thin=thick}).
The solved $B_{\rm tot}$ and $K_{e}$ are written 
as functions of 
$\theta_{\rm obs}$, $\gamma_{e,\rm min}$,
$\nu_{\rm ssa}$,  and the observed flux  at $\nu=\nu_{\rm ssa}$.

Lastly, we further impose total jet power 
constraint not to overproduce Poynting- or kinetic-power  
shown in Eq.~(\ref{eq:energetics}).
This constraint can partially
exclude larger  value of $\theta_{\rm obs}$.
Then, we can determine $B_{\rm tot}$ and $U_{e}/U_{\rm B}$
of M87 consistently.

\subsection{Assumptions}

Following assumptions are adopted in this work:

\begin{itemize}

\item
We assume uniform and isotropic distribution
of relativistic electrons and magnetic fields 
in the emission region.
For M87, polarized flux does not seem very large.
Therefore, we assume isotropic tangled
magnetic field in this work.
Hereafter, we denote $B$ as the magnetic field strength
perpendicular to the direction of electron motion.
Then, the total field strength is 
\begin{eqnarray}
B_{\rm tot}=\sqrt{3} B  .
\end{eqnarray}
Hereafter, we define $U_{B}\equiv B_{\rm tot}^2/8\pi$.

\item 
We assume the emission region is spherical
with its radius $R$ measured in the comoving frame. 
The radius is defined as
\begin{eqnarray}
2R=\theta_{\rm obs}D_{A} ,
\end{eqnarray}
where $D_{A}=D_{L}/(1+z)^{2}$ is 
the angular diameter distance 
to a source  (e.g., Weinberg 1972).
Because M87 is 
the very low redshift source, 
we only use $D_{A}$ throughout this paper.
There might be a slight difference between
$\theta_{\rm FWHM}$ and $\theta_{\rm obs}$.
VLBI measured $\theta_{\rm FWHM}$ is conventionally 
treated as $\theta_{\rm obs}=\theta_{\rm FWHM}$,
while Marscher (1983)
pointed out a deviation expressed as 
$\theta_{\rm obs}\approx 1.8\theta_{\rm FWHM}$ 
which is caused by a forcible fitting of Gaussians 
to a non-Gaussian component.
In this work, we introduce a factor $A$ defined as 
$\theta_{\rm obs}\equiv A \theta_{\rm FWHM}$ and  
$1\le A \le 1.8$ is assumed.

\end{itemize}

We stress that the uniform and isotropic sphere 
model is a first step simplification 
and the realistic jet base probably contains 
more complicated geometry and nonuniform distributions
in magnetic field and electron density. 
We will investigate such complicated 
cases in the future.

\subsection{Synchrotron emissions and absorptions}

In order to obtain explicit expression of 
$B$ and $K_{e}$ in terms of  $\theta_{\rm obs}$, 
$\nu_{\rm ssa, obs}$, and $S_{\nu_{\rm ssa},\rm obs}$,
here we briefly review synchrotron emissions and absorptions.
At the radio core, 
$\tau_{\nu_{\rm ssa}}$ becomes an order of unity
at $\nu=\nu_{\rm ssa}$;
\begin{eqnarray}\label{eq:tau}
\tau_{\nu_{\rm ssa}}= 2 \alpha_{\nu_{\rm ssa}} R ,
\end{eqnarray}
where 
$\tau_{\nu_{\rm ssa}}$ and
$\alpha_{\nu_{\rm ssa}}$ 
are
the optical depth for SSA and
the absorption coefficient for SSA, respectively.
We impose
that optically thin emission formula is still applicable
at $\nu_{\rm ssa}$:
\begin{eqnarray}\label{eq:thin=thick}
\frac{4\pi}{3} R^{3} \epsilon_{\nu_{\rm ssa}} 
&=& 4\pi R^{2} S_{\nu_{\rm ssa}}   ,
\end{eqnarray}
where 
$\epsilon_{\nu_{\rm ssa}}$ and
$S_{\nu_{\rm ssa}}$ are
he emissivity and flux per unit frequency,
respectively.
Combining Eq.~\ref{eq:thin=thick} and the approximation
of $\tau_{\nu_{\rm ssa}}=1$, 
we can solve $B$ and $K_{e}$.
This derivation is much simpler than  
previous studies of Marscher (1983) 
and Hirotani (2005) (hereafter H05).
We will compare 
the  derived $B_{\rm tot}$ in this work, 
Marscher (1983) and H05, and they will
coincide with each other within a small difference 
in the range of $2.5\le p \le 3.5$.

Next, let us break down relevant physical quantities.
The term $K_{e}$, the normalization factor of 
electron number density distribution $n_{e}(\gamma)$, 
is defined as (e.g., Eq.3.26 in GS65)
\begin{eqnarray}
n_{e}(E_{e})dE_{e} &=&
K_{e}E_{e}^{-p} dE_{e} 
\quad (E_{e,\rm min}\le E_{e} \le E_{e,\rm max})  , \nonumber \\
&=&
\frac{K_{e}}{(m_{e}c^{2})^{p-1}} \gamma_{e}^{-p}d\gamma_{e} 
\quad (\gamma_{e,\rm min}\le \gamma_{e} \le \gamma_{e,\rm max}), \nonumber \\  
\end{eqnarray}
where 
$E_{e}=\gamma_{e}m_{e}c^{2}$
$p=2\alpha+1$,
$E_{e,{\rm min}}=\gamma_{e,{\rm min}}m_{e}c^{2}$, and
$E_{e,{\rm max}}=\gamma_{e,{\rm max}}m_{e}c^{2}$
are
the electron energy,
the spectral index,
minimum energy, and
maximum energy of 
relativistic (non-thermal) electrons, respectively.
Let us further review optically thin
synchrotron emissions.
The maximum in the spectrum of synchrotron radiation
from an electron occurs at the frequency:
(Eq. 2.23 in GS65)
\begin{eqnarray}\label{eq:nu_syn}
\nu_{\rm syn}=1.2\times 10^{6}B\gamma_{e}^{2}   .
\end{eqnarray}
Synchrotron self-absorption coefficient measured
in the comoving frame is given by
(Eqs.~4.18 and 4.19 in GS65; Eq. 6.53 in RL79)
\begin{eqnarray}
\alpha_{\nu}&=&
\frac{\sqrt{3}e^{3}}{8\pi m_{e}}
\left(\frac{3e}{2\pi m_{e}^{3}c^{5}}\right)^{p/2}
c_{1}(p) \nonumber \\
&\times&
K_{e} B^{(p+2)/2}
\nu^{-(p+4)/2},
\end{eqnarray}
where the numerical coefficient $c_{1}(p)$
is expressed by using the gamma-functions as follows;
$c_{1}(p)=
\Gamma[(3p+2)/12]
\Gamma[(3p+22)/12]$.
For convenience, we define 
$\alpha_{\nu}=X_{1}c_{1}(p)B^{(p+2)/2}K_{e}\nu^{-(p+4)/2}$.

Optically thin synchrotron emissivity
per unit frequency $\epsilon_{\nu}$ 
from uniform emitting region is given by
(Eqs. 4.59 and 4.60 in BG70;
see also Eqs. 3.28, 3.31 and 3.32 in GS65)
%
\begin{eqnarray}
\epsilon_{\nu} &=& 
4\pi\frac{\sqrt{3}e^{3}}{8 \sqrt{\pi} m_{e}c^{2}}
\left(\frac{3e}{2\pi  m_{e}^{3}c^{5}}\right)^{(p-1)/2} 
c_{2}(p) \nonumber \\ 
&\times& K_{e}B^{(p+1)/2} \nu^{-(p-1)/2}   ,
\end{eqnarray}
where the numerical coefficient is 
$c_{2}(p)=
\Gamma[(3p+19)/12]
\Gamma[(3p-1)/12)]
\Gamma[(p+5)/4)]/\Gamma[(p+7)/4)]
/(p+1)$.
For convenience, we define 
$\epsilon_{\nu} \equiv 4\pi X_{2} c_{2}(p) B^{(p+1)/2}K_{e}
\nu^{-(p-1)/2}$.


\subsection{Relations between quantities 
measured in source and observer frames}

Let us summarize the Lorentz transformations
and cosmological effect using the 
Doppler factor ($\delta\equiv 1/(\Gamma(1-\cos\theta_{\rm LOS})$)
where $\theta_{\rm LOS}$ is the angle
between the jet and our line-of-sight)
and the redshift ($z$).
Hereafter, we put subscript (obs) for quantities
measured at observer frame.
%
\begin{eqnarray}\label{eq:nu_dopp}
\nu_{\rm obs} = \nu \frac{\delta}{1+z}.
\end{eqnarray}
The observed flux from an optically
thin source at a large distance is given  
by (Eq. 1.13 in RL79; Eqs.~(7) in H05;
see also Eq.~C4 in Begelman et al. 1984):
\begin{eqnarray}
S_{\nu_{\rm obs},\rm obs}
&=&
\left(\frac{\delta}{1+z}\right)^{3}
S_{\nu} 
\left(\frac{\theta_{\rm obs}}{2}\right)^{2}  .
\end{eqnarray}
%

\subsection{Obtained $B$ and $K_{e}$}

%

Combining the above shown relations,
we finally obtain 
\begin{eqnarray}\label{eq:B}
B&=& b(p)
\left(\frac{\nu_{\rm ssa, obs}}{1~{\rm GHz}}\right)^{5}
\left(\frac{\theta_{\rm obs}}{1~{\rm mas}}\right)^{4}
\left(\frac{S_{\nu_{\rm ssa},\rm obs}}{1~{\rm Jy}}\right)^{-2} \nonumber \\
&\times& \left(\frac{\delta}{1+z}\right)    ,
\end{eqnarray}
where the numerical value of 
$b(p)=[(2\times 3)/(4\pi)]^{2}(c_{2}(p)X_{2}/c_{1}(p)X_{1})^{2}$ 
are shown in Table~\ref{table:coefficient}.
In the Table, 
we also note the values the obtained $b(p)$ with the ones in 
Marscher (1983) and H05.
From this, we see that
the derived $B_{\rm tot}$ in this work coincide 
with each other within the small difference.

Inserting Eq.~(\ref{eq:B}) into Eq.~(\ref{eq:tau})
or  Eq.~(\ref{eq:thin=thick}),
we then  obtain $K_{e}$ as
\begin{eqnarray}
K_{e} &=& k(p)
\left(\frac{D_{\rm A}}{1~{\rm Gpc}}\right)^{-1}
\left(\frac{\nu_{\rm ssa,obs}}{1~{\rm GHz}}\right)^{-2p-3}
\left(\frac{\theta_{\rm obs}}{1~{\rm mas}}\right)^{-2p-5} \nonumber \\
&\times&
\left(\frac{S_{\nu_{\rm ssa},\rm obs}}{1~{\rm Jy}}\right)^{p+2}
\left(\frac{\delta}{1+z}\right)^{-p-3},
\end{eqnarray}
where $k(p) = b(p)^{(-p-2)/2} 
X_{1}^{-1}c_{1}(p)^{-1}
(D_{\rm A}/1~{\rm Gpc})^{-1} \nu_{\rm ssa,obs}^{(p+4)/2}$.  
The cgs units of $K_{e}$ and $k(p)$
depend on $p$:
erg$^{p-1}$cm$^{-3}$.
The numerical values of $k(p)$ are summarized in Table~1 and
they are similar to the ones in
Marscher (1983) which are
$k(2.5)=1.2\times 10^{-2}$ and 
$k(3.0)=0.59 \times 10^{-3}$.
Using the obtained $K_{e}$, we can evaluate
$U_{e}$ as
\begin{eqnarray}
U_{e} &=& \int^{E_{e,\rm max}}_{E_{e,\rm min}}  
E_{e}n_{e}(E_{e})dE_{e}  \nonumber \\
&=& \frac{K_{e}E_{e,\rm min}^{-p+2}}{p-2}
\quad ({\rm for} \quad p>2)  .
\end{eqnarray}
Then, we can obtain the ratio
$U_{e}/U_{B}$ explicitly as
\begin{eqnarray}\label{eq:ueub}
\frac{U_{e}}{U_{B}}&=& 
\frac{8\pi}{3b^{2}(p)} 
\frac{k(p)E_{e,\rm min}^{-p+2}}{(p-2)} 
\left(\frac{D_{\rm A}}{1~{\rm Gpc}}\right)^{-1}
\left(\frac{\nu_{\rm ssa,obs}}{1~{\rm GHz}}\right)^{-2p-13}
\nonumber \\
&\times&
\left(\frac{\theta_{\rm obs}}{1~{\rm mas}}\right)^{-2p-13} 
\left(\frac{S_{\nu_{\rm ssa},\rm obs}}{1~{\rm Jy}}\right)^{p+6}
\left(\frac{\delta}{1+z}\right)^{-p-5}  \nonumber \\
&& ({\rm for} \quad p>2)  .
\end{eqnarray}
From this, 
we find that $\nu_{\rm ssa,obs}$ and
$\theta_{\rm obs}$ have the same dependence on $p$.
Using this relation, we can estimate $U_{e}/U_{B}$ 
without minimum energy (equipartition $B$ field) assumption.
It is clear that 
the measurement of $\theta_{\rm obs}$ is crucial
for determining $U_{e}/U_{B}$.
We argue details on it in the next subsection.
It is also evident that
a careful treatment of $\gamma_{e,\rm min}$
is crucial for determining $U_{e}/U_{B}$
(Kino et al. 2002; Kino and Takahara 2004).

\section{Application to M87}

Based on recent VLBA observations of M87 at 43~GHz,
we derive $U_{e}/U_{\rm B}$ at the base of 
M87 jet.
Here, we set typical index of electrons as $p=3.0$
(Doi et al. 2013).

\subsection{Electrons emitting 43~GHz radio waves}

First, let us constrain on $\gamma_{e,\rm min}$.
At least, $\gamma_{e,\rm min}$ 
should be smaller than  the Lorentz factor of electrons
radiating synchrotron emission at 43~GHz.
Therefore, 
minimum Lorentz factor of electrons is constrained as
\begin{eqnarray}\label{eq:gmin}
\gamma_{e,\rm min} &\le& 2.0\times 10^{2}
\left(\frac{\nu_{\rm ssa,obs}}{43~{\rm GHz}}\right)^{-2}  
\left(\frac{\theta_{\rm obs}}{0.11~{\rm mas}}\right)^{-2}  \nonumber \\
&\times& \left(\frac{S_{\nu_{\rm ssa},\rm obs}}{0.7~{\rm Jy}}\right)^{1}
\left(\frac{\delta}{1+z}\right)^{-1} .
\end{eqnarray}
where we use Eqs. (\ref{eq:nu_syn}) and (\ref{eq:nu_dopp}).

\subsection{Normalized physical quantities}

For convenience, we rewrite
above quantities to normalized quantitates
associated with the observed 43~GHz core. 
$B$ field inside the 43~GHz core is estimated as
\begin{eqnarray}\label{eq:B_M87}
B_{\rm tot}&=& 1.5~{\rm G}~
\left(\frac{\nu_{\rm ssa, obs}}{43~{\rm GHz}}\right)^{5}
\left(\frac{\theta_{\rm obs}}{0.11~{\rm mas}}\right)^{4}
\left(\frac{S_{\nu_{\rm ssa},\rm obs}}{0.7~{\rm Jy}}\right)^{-2} \nonumber \\
&\times& \left(\frac{\delta}{1+z}\right)   .
\end{eqnarray}
Here we use $ 0.11~{\rm mas}$ as a
normalization of $\theta_{\rm FWHM}$. 
As mentioned in the Introduction,
$\nu_{\rm ssa, obs}=43~{\rm GHz}$ holds 
at the 43~GHz core surface, because of  the 
clear detection of the core shift phenomena in H11.
As for $K_{e}$, we obtain
\begin{eqnarray}
K_{e} &=& 
1.6\times 10^{-6}~{\rm erg^{2}~cm^{-3}}~
\left(\frac{\nu_{\rm ssa,obs}}{43~{\rm GHz}}\right)^{-9}
\left(\frac{\theta_{\rm obs}}{0.11~{\rm mas}}\right)^{-11} \nonumber \\
&\times&
\left(\frac{S_{\nu_{\rm ssa},\rm obs}}{0.7~{\rm Jy}}\right)^{5}
\left(\frac{\delta}{1+z}\right)^{-6}
\left(\frac{D_{\rm A}}{16.7~{\rm Mpc}}\right)^{-1} .
\end{eqnarray}
Then, we finally obtain $U_{e}/U_{B}$ as
\begin{eqnarray}\label{eq:UeUb_M87}
\frac{U_{e}}{U_{B}}&=&
2.2~ 
\left(\frac{\nu_{\rm ssa,obs}}{43~{\rm GHz}}\right)^{-19}
\left(\frac{\theta_{\rm obs}}{0.11~{\rm mas}}\right)^{-19} 
\nonumber \\
&\times&
\left(\frac{S_{\nu_{\rm ssa},\rm obs}}{0.7~{\rm Jy}}\right)^{9}
\left(\frac{\delta}{1+z}\right)^{-8}
\left(\frac{\gamma_{e,\rm min}}{10}\right)^{-1}  \nonumber \\
&\times&
\left(\frac{D_{\rm A}}{16.7~{\rm Mpc}}\right)^{-1} .
\end{eqnarray}
This typical $U_{e}/U_{B}$ apparently
shows the order of unity
but it has strong dependences 
on $\theta_{\rm obs}$ and $\nu_{\rm ssa}$. 
Regarding $\nu_{\rm ssa,obs}$, 
an uncertainty 
only comes from the bandwidth. 
In our VLBA observation,
the bandwidth is $128~{\rm MHz}$
with its central frequency 43.212~GHz. 
Therefore, it causes only a very small
uncertainty $\sim (43.276/43.148)^{19}=1.06$.
The accuracy of flux calibration of VLBA  can
be conservatively estimated as $10~\%$.
An intrinsic flux of the radio core at 43~GHz
also fluctuates with an order of $10~\%$ 
during a quiescent phase 
(e.g., Acciari et al. 2009; Hada et al. 2012).
Therefore, the flux term also causes
an uncertainty of $\sim 1.21^{9}=5.6$.
The angular size $\theta_{\rm obs}$ and
$\gamma_{e,\rm min}$
have much larger ambiguities than those 
evaluated above
and we derive $U_{e}/U_{\rm B}$
by taking into  account of these ambiguities
in the next sub-section.
At the same time, we again emphasize that 
$\theta_{\rm obs}$ obtained in H13 can be bearable
for the estimate of $U_{e}/U_{\rm B}$ in spite of  
such strong dependence on $\theta_{\rm obs}$.

\subsection{On $\theta_{\rm obs}$, $p$, and $L_{\rm jet}$}

The most important 
quantity for
the estimate of $U_{e}/U_{\rm B}$ is $\theta_{\rm obs}$. 
Based on 
VLBA observation data with sufficiently
good qualities, here we set
\begin{eqnarray}\label{eq:theta_M87}
0.11~{\rm mas}\le  \theta_{\rm obs}\le  0.20~{\rm mas} ,  
\end{eqnarray}
where we use the average value
$\theta_{\rm FWHM}=0.11~{\rm mas}$ from H13 
and maximum of $\theta_{\rm obs}$ is 
$0.11~{\rm mas}\times 1.8=0.198~{\rm mas}$.
We note that the measured
core's FWHM overlaps with the 
measured width of the jet (length between
the jet limb-structure) in H13.
Therefore, we consider $A \approx 1$ more likely
for the M87 jet base.
From Eq. (\ref{eq:gmin}),
the maximal value of $\gamma_{e,\rm min}$ 
is given by
$\sim 2\times 10^{2}$
when $\theta_{\rm obs}=0.11~{\rm mas}$.


Regarding the value of $p=2\alpha+1$, 
a simultaneous observation of the spectrum measurement
at sub-mm wavelength range is crucial, 
since most of the observed fluxes at sub-mm range
come from the innermost part of the jet.
It has been indeed measured by Doi et al. (2013)
by conducting
a quasi-simultaneous multi-frequency observation
with the Atacama Large 
Millimeter/submillimeter Array (ALMA) observation 
(in cycle 0 slot)
and it is robust that $\alpha >0.5$  at $>200~{\rm GHz}$ 
where 
synchrotron emission becomes optically-thin against SSA.
Maximally taking uncertainties into account, 
we set the allowed range of $p$ as
\begin{eqnarray}
2.5\le  p \le  3.5,  
\end{eqnarray}
in this work.

We further impose the condition that
time-averaged total jet power  ($L_{\rm jet}$)
inferred from its large-scale jet properties
should not be exceeded 
by 
the kinetic power of relativistic electrons  ($L_{e}$)
and Poynting power  ($L_{\rm poy}$) at the 43~GHz core
\begin{eqnarray}\label{eq:energetics}
L_{\rm jet} &\ge&  \max[L_{\rm poy}, L_{e}] ,  \nonumber \\
L_{e} &=& \frac{4\pi}{3} \Gamma^{2} \beta R^{2} c U_{e} , \nonumber \\
L_{\rm poy} &=& \frac{4\pi}{3} \Gamma^{2} \beta R^{2} c U_{B} , 
\end{eqnarray}
where $L_{\rm jet}$ at large-scale 
is estimated maximally 
a few $\times 10^{44}~{\rm erg~s^{-1}}$
(e.g., 
Reynolds et al. 1996;
Bicknell and Begelman 1996;
Owen et al. 2000; 
Stawarz et al. 2006;
Rieger and Aharonian 2012).
Hereafter, we conservatively assume $\Gamma\beta=1$ 
and a slight deviation from this 
does not influence the main results in this work.
Regarding $L_{\rm jet}$ in the M87 jet, we set 
\begin{eqnarray}\label{eq:Ljet}
1 \times 10^{44}~{\rm erg~s^{-1}}\le 
L_{\rm jet}\le 5 \times 10^{44}~{\rm erg~s^{-1}}  .
\end{eqnarray}
%
Here we include an uncertainty
due to the deviation from time-averaged 
$L_{\rm jet}$ at large-scale which may attribute
to flaring phenomena at the jet base.
X-ray light curve at the M87 core 
over 10 years showed a flux variation
by a factor of several except for 
exceptionally high X-ray flux during 
giant VHE flares happened in 2008 and 2010
(Fig.~1 in Abramowski et al. 2012).
Based on it, 
we set the largest jet kinetic power case as
$L_{\rm jet}
=5 \times 10^{44}~{\rm erg~s^{-1}}$.



\subsubsection{On jet speed}

Jet speed in the vicinity of M87's 
central black hole is quite an issue.
Ly et al. (2007) and Kovalev et al. (2007)
show sub-luminal speed
proper motions of the M87 jet base.
The recent study by Asada et al. (2014)
also support it.
Hada (2013b) also explores the proper
motion near the jet base with 
the VERA (VLBI Exploration of Radio Astrometry) array.
The VERA observation has been partly 
performed in the GENJI programme 
(Gamma-ray Emitting Notable
AGN Monitoring with Japanese VLBI) aiming 
for densely-sampled monitoring of bright AGN jets
(see Nagai et al. 2013 for details)
and the observational data obtained by VERA
also shows  a sub-luminal motion at the jet base.
Furthermore,
Acciari et al. (2009) report that
the 43~GHz core is stationary within $\sim 6~R_{\rm s}$
based on their 
phase-reference observation at 43~GHz.
Therefore, currently there is no clear observational
support of  super-luminal motion within the 43~GHz radio core.
The brightness temperature 
$T_{b}=
\frac{1+z}{\delta}
\frac{S_{\nu_{\rm obs},\rm obs}c^{2}}{2\pi k \nu_{\rm obs}^{2} 
(\theta_{\rm obs}/2)^{2}}$
(e.g., 
L{\"a}hteenm{\"a}ki et al. 1999;
Doi et al. 2006)
of the 43~GHz radio core
is evaluated as 
\begin{eqnarray}
T_{b} \sim  6 \times 10^{10}~{\rm K}
\left(\frac{S_{\nu_{\rm ssa}, \rm obs}}{0.7~{\rm Jy}}\right)
\left(\frac{\theta_{\rm obs}}{0.11~{\rm mas}}\right)^{-2}  ,
\end{eqnarray}
which is below the critical temperature $\sim 10^{11}~{\rm K}$ 
limited by inverse-Compton catastrophe process
(Kellermann and Pauliny-Toth 1969).
Because of these two reasons, we assume 
$\delta\approx 1$ throughout this paper.

\section{Results}

Here, we examine the three cases of 
electron indices as
$p=$2.5, 3.0, and 3.5 against the two cases of 
the jet power as 
$L_{\rm jet}=1\times 10^{44}~{\rm erg~s^{-1}}$ and
$L_{\rm jet}=5\times 10^{44}~{\rm erg~s^{-1}}$.

\subsection{Allowed B strength}

First of all, it should be noted that $B$ is primarily 
determined by a value of $\theta_{\rm obs}$
since $\nu_{\rm ssa}$ is exactly identical to
the observing frequency.
By combining Eqs.~(\ref{eq:B_M87})
~(\ref{eq:energetics}), and
(\ref{eq:Ljet}),
we obtain the allowed range of magnetic field strength
in the 43~GHz core.
We summarize the
obtained maximum and minimum values of $B_{\rm tot}$
in Table~\ref{table:B}.
An upper limit of $B$ is governed by
the constraint of $L_{\rm jet}\ge L_{\rm poy}$.
From Eqs. (\ref{eq:B}) and (\ref{eq:energetics}), 
it is clear that $L_{\rm poy}$ behaves as
\begin{eqnarray}\label{eq:energetics2poy}
L_{\rm poy}&\propto& \theta_{\rm obs}^{10}
            \propto  B_{\rm tot}^{5/2}    .
\end{eqnarray}
%
apart from a weak dependence on
$p$ originating in $b(p)$.
We thus obtain the allowed range
$ 1~{\rm G}\le B_{\rm tot} \le 15~{\rm G}$
in the 43~GHz core.
This is a robust constraint 
on the M87 core's B strength.

\subsection{Allowed $U_{e}/U_{B}$ with $p=3.0$}

In Fig.~\ref{fig:ueub_5d44_p30}, 
we show the allowed 
region in $\gamma_{e,\rm min}$ and $B_{\rm tot}$ plane
(the red-colored boxed region)
and the corresponding $\log (U_{e}/U_{B})$ values
with 
$L_{\rm jet}=5\times 10^{44}~{\rm erg~s^{-1}}$
and $p=3.0$ 
which is based on the power law index 
measured by ALMA (Doi et al. 2013).
The larger $\gamma_{e,\rm min}$ leads to
smaller $\log (U_{e}/U_{B})$ because $U_{e}$
becomes smaller for  larger $\gamma_{e,\rm min}$.
Similar to the aforementioned 
$B_{\rm tot}$ constraints,
the lower limit of $U_{e}/U_{B}$ is bounded by
$L_{\rm jet}=L_{\rm poy}$,
while the upper limit
is governed by the lowest value of $B_{\rm tot}$.
From this, we conclude that 
both $U_{e}/U_{B}>1$
and $U_{e}/U_{B}<1$ can be possible.
We again stress that the field strength $B_{\rm tot}$ 
has one-to-one correspondence to $\theta_{\rm obs}$.
In other words,
acculate determination of $\theta_{\rm obs}$
is definitely important for the estimate of  $U_{e}/U_{B}$.
By the energetic constraint shown in Eq.~(\ref{eq:energetics}),
the  maximum $\theta_{\rm obs}$ becomes smaller than
$0.20~{\rm mas}$. In this case, we obtain
\begin{eqnarray}
0.11~{\rm mas} \le \theta_{\rm obs} \le 0.19~{\rm mas}  .
\end{eqnarray}
The maximum $\theta_{\rm obs}$ is derived from 
$L_{\rm jet}=L_{\rm poy}=5\times 10^{44}~{\rm erg~s^{-1}}$.
Then, the factor
$(0.19/0.11)^{18}\sim 2 \times 10^{4}$
makes the allowed $U_{e}/U_{B}$ range broaden.
Independent of this factor, $\gamma_{e,\rm min}$
has uncertainty about the factor of $2\times 10^{2}$.
These factors govern
the overall allowed $U_{e}/U_{B}$ range
of the order of a few times $10^{6}$
which is presented in Table 3.
Additionally, we note that the right-top part 
is dropped out according to Eq.~(\ref{eq:gmin}). 
This changes minimum values 
of $U_{e}/U_{\rm B}$ by a
factor of a few.



Additionally,
we show the allowed $\log (U_{e}/U_{B})$ and $B$
with 
$L_{\rm jet}=1\times 10^{44}~{\rm erg~s^{-1}}$
and $p=3.0$ in Tables 2 and 3.
Compared with the case in Fig.~\ref{fig:ueub_5d44_p30}, 
the upper limit of $L_{\rm poy}$ becomes 
smaller.
Then the allowed  $\theta_{\rm obs}$ becomes
\begin{eqnarray}
0.11~{\rm mas} \le \theta_{\rm obs} \le 0.16~{\rm mas}  .
\end{eqnarray}
The maximum $\theta_{\rm obs}$ is also derived from 
$L_{\rm jet}=L_{\rm poy}=1\times 10^{44}~{\rm erg~s^{-1}}$.
The decrease of the  
maximum $\theta_{\rm obs}$ value
leads to the increase of
$U_{e}/U_{\rm B}$ correspondingly.

\subsection{Allowed $U_{e}/U_{B}$ with $p=3.5$}

In Fig.~\ref{fig:ueub_5d44_p35}, 
we show the allowed 
region in  $\gamma_{e,\rm min}$ and $B_{\rm tot}$ plane
(the red-colored boxed region)
and the corresponding $\log (U_{e}/U_{B})$ values
with 
$L_{\rm jet}=5\times 10^{44}~{\rm erg~s^{-1}}$
and $p=3.5$.
Compared with the case with $p=3.0$,
the $U_{e}/U_{\rm B}>1$ region increases
in the allowed parameter range 
according to the relation of
$U_{e}/U_{\rm B} \propto 
 \theta_{\rm obs}^{-2p-13}
 \gamma_{e,\rm min}^{-p+2}$ .
The allowed $\theta_{\rm obs}$ in this case is 
\begin{eqnarray}\label{eq:allowed_theta}
0.11~{\rm mas} \le \theta_{\rm obs} \le 0.20~{\rm mas}  ,
\end{eqnarray}
which remains the same as Eq.~(\ref{eq:theta_M87})
because both $L_{e}$ and $L_{\rm poy}$ do not exceed
$L_{\rm jet}=5\times 10^{44}~{\rm erg~s^{-1}}$ in this case.
The $\theta_{\rm obs}$
factor leads to $(0.196/0.11)^{20}\approx 1 \times 10^{5}$
uncertainty while $\gamma_{e,\rm min}$ factor 
has $\sim 2\times 10^{2}$ uncertainty. Therefore, 
the allowed $U_{e}/U_{\rm B}$ in this case
has a few times $10^{7}$ of uncertainty which 
is shown in Table 3.
The left-bottom part is slightly 
deficit because of  $L_{e}>L_{\rm jet}$.

In Tables 2 and 3,
we show the allowed $\log (U_{e}/U_{B})$ and $B$
with $L_{\rm jet}=1\times 10^{44}~{\rm erg~s^{-1}}$
and $p=3.5$.
In this case, the allowed $\theta_{\rm obs}$ is
\begin{eqnarray}
0.11~{\rm mas} \le \theta_{\rm obs} \le 0.18~{\rm mas}  .
\end{eqnarray}
The relation of
$L_{\rm jet}=L_{\rm poy}=1\times 10^{44}~{\rm erg~s^{-1}}$
leads to the value of maximum $\theta_{\rm obs}$.
This upper and lower $U_{e}/U_{B}$ 
are governed in the same way as in 
in Fig.~\ref{fig:ueub_5d44_p35}.

\subsubsection{Allowed $U_{e}/U_{B}$ with $p=2.5$}

In Fig.~\ref{fig:ueub_5d44_p25}, 
we show the allowed 
region in  $\gamma_{e,\rm min}$ and $B_{\rm tot}$ plane
(the red-colored boxed region)
and the corresponding $\log (U_{e}/U_{B})$ values
with $L_{\rm jet}=5\times 10^{44}~{\rm erg~s^{-1}}$
and $p=2.5$.
In this case, the allowed $\theta_{\rm obs}$ is
\begin{eqnarray}
0.11~{\rm mas} \le \theta_{\rm obs} \le 0.17~{\rm mas}  .
\end{eqnarray}
The relation of
$L_{\rm jet}=L_{\rm poy}=5\times 10^{44}~{\rm erg~s^{-1}}$
determines the maximum $\theta_{\rm obs}$.
The allowed $B_{\rm tot}$ is in the narrow
range of 
$2.5~{\rm G}\le B_{\rm tot} \le 14.7~{\rm G}$.
It should be stressed that this case shows the
magnetic field energy dominande  in all of 
the allowed $B_{\rm tot}$ $\gamma_{e,\rm min}$
ranges.

In Tables 2 and 3,
we show the allowed $\log (U_{e}/U_{B})$ and $B$
with $L_{\rm jet}=1 \times 10^{44}~{\rm erg~s^{-1}}$
and $p=2.5$.
Basic behavior is similar to the case 
shown in Fig.~\ref{fig:ueub_5d44_p25}.
In this case, the allowed $\theta_{\rm obs}$ is
\begin{eqnarray}
0.11~{\rm mas} \le \theta_{\rm obs} \le 0.15~{\rm mas}  .
\end{eqnarray}
The relation of
$L_{\rm jet}=L_{\rm poy}=1\times 10^{44}~{\rm erg~s^{-1}}$
determines the maximum $\theta_{\rm obs}$.
Corresponding to the narrow allowed range of $\theta_{\rm obs}$,
the allowed field strength resides in the narrow range of
$2.5~{\rm G} \le B_{\rm tot} \le 7.7~{\rm G}$.

\section{Summary and discussions}

Based on VLBA observation data at 43~GHz,
we explore $U_{e}/U_{B}$ at the base of the M87 jet.
We apply the standard theory of synchrotron radiation
to the 43~GHz radio core
together with the assumption of 
a simple uniform sphere geometry.
We impose the condition
that the Poynting and 
relativistic electron kinetic power should
be smaller than the total power of the jet. 
Obtained values 
of $B_{\rm tot}$ and $U_{e}/U_{\rm B}$
are summarized in
Tables \ref{table:B} and \ref{table:UeUb}
and we find the followings;

\begin{itemize}

\item

We obtain the allowed range of magnetic field 
strength in the 43~GHz core as
$1~{\rm G} \le B_{\rm tot} \le 15~{\rm G}$ in the observed 
radio core at 43~GHz with its diameter 
$0.11\--0.20~{\rm mas}$ $(15.5\-- 28.2 ~R_{\rm s})$.
Our estimate of $B$  is basically close 
to the previous estimate in  the literature 
(e.g., Neronov and Aharonian 2007),
although fewer assumptions have been 
made in this work.
We add to note that even
if $\delta$ of the 43~GHz core becomes larger than unity, 
the field strength only changes according to
$B_{\rm tot}\propto \delta$.

It is worth to compare these values with 
independently estimated $B_{\rm tot}$ in previous works
more carefully.
Abdo et al. (2009) has estimated
Poynting power and kinetic power of the jet
by the model fitting 
of the observed broad band spectrum
and derive 
$B_{\rm tot}=0.055~{\rm G}$
with $R=1.4\times 10^{16}~{\rm cm}=0.058~{\rm mas}$,
although they do not properly include SSA effect.
Acciari et al. (2009) 
predict field strength
$B_{\rm tot}\sim 0.5~{\rm G}$
based on the synchrotron cooling 
argument.
Since smaller values of $B_{\rm tot}$ lead 
to smaller $\theta_{\rm obs}$, if we assume  
$\theta_{\rm obs,min}$ by a factor of $\sim 3$
than the true $\theta_{\rm obs}=0.11~{\rm mas}$,
the predicted $B_{\rm tot}$ lies between 0.05 and 0.5 gauss
which seems to be in a good agreement with previous work.
However, 
for such a small core, electron 
kinetic power much exceeds the observed jet power.

Our result excludes a strong magnetic field
such as $B_{\rm tot}\sim 10^{3-4}~{\rm G}$ which is
frequently assumed in previous works in order to 
activate Blandford-Znajek process 
(Blandford and Znajek 1977;
Thorne et al. 1986;
Boldt \& Loewenstein 2000).
Although M87 has been a prime target for testing
relativistic MHD jet simulation studies powered by black-hole spin
energy, our result throw out the caveat that
the maximum $B_{\rm tot}$, one of the critical parameters
in relativistic MHD jets model, 
$B_{\rm tot}$ should be smaller 
than $\sim 15$~G for M87.

\item

We obtain the allowed region of $U_{e}/U_{B}$ 
in the allowed $\theta_{\rm obs}$ and $\gamma_{e,\rm min}$
plane. 
The resultant $U_{e}/U_{B}$ contains both the 
region of  $U_{e}/U_{B}>1$ and  $U_{e}/U_{B}<1$.
It is found that the allowed range is 
$1\times 10^{-5}\le U_{e}/U_{B}\le 6\times 10^{2}$.
The uncertainty of $U_{e}/U_{B}$  is caused 
by the
strong dependence  on $\theta_{\rm obs}$ and $\gamma_{e,\rm min}$.
Our result  gives an important constraint 
against relativistic MHD models 
in which they postulate 
very large $U_{\rm B}/U_{e}$ at a jet-base 
(e.g., 
Vlahakis and Konigel 2003;
Komissarow et al. 2007, 2009;
Tchekhovskoy et al. 2011).
To realize sufficiently
magnetic dominated jet such 
as $U_{\rm B}/U_{e}\sim 10^{3-4}$, 
relatively large
$\gamma_{e,\rm min}$ of the 
order of $\sim 10^{2}$ 
and a relatively large $\theta_{\rm obs}$ are
required.
Thus, the obtained $U_{e}/U_{\rm B}$ 
in this work gives a new constraint 
on the initial conditions 
in relativistic MHD models.

\end{itemize}

Last, we shortly note  key future works.

\begin{itemize}

\item 
Observationally,
it is crucial to obtain
resolved images of the radio cores at 43~GHz
with space/sub-mm VLBI which would clarify 
whether there is a sub-structure or not 
inside $\sim 16$~Rs scale at the M87 jet base.
Towards this observational final goal,
as a first step,
it is important to explore physical relations 
between the results of the  present work and 
observational data at higher frequencies such as 86~GHz 
and 230~GHz
(e.g.,
Krichbaum et al. 2005;
Krichbaum et al. 2006;
Doeleman et al. 2012).
Indeed, 
we  conduct a new observation of M87 
with VLBA and the Green Bank Telescope
at 86~GHz and we will explore this issue
using the new data.
Space-VLBI program also
could play a key role since lower frequency observation
can attain higher dynamic range images with a high 
resolution 
(e.g., Dodson et al. 2006;
Asada et al. 2009; 
Takahashi and Mineshige 2010;
Dodson et al. 2013).
If more compact regions inside the
0.11mas region are found by space-VLBI in the future,
then $U_{e}/U_{\rm B}$ in the compact regions
are larger than the ones shown in the present work.

\item

Theoretically, we leave
following issues as our future work.
(1)
Constraining plasma composition
(i.e, electron/proton ratio) is
one of the most important issue in AGN jet physics
(Reynolds et al. 1996; Kino et al. 2012)
and we will study it in the future.
Roughly saying, inclusion of proton powers ($L_{p}$)
will simply reduce the upper limit of 
$B_{\rm tot}$ because 
$L_{\rm jet} \approx L_{e}+L_{p}\approx L_{\rm poy}$
would hold.
(2)
On $\sim 10~R_{\rm s}$ scale,
general relativistic (GR) effects can be important and
they will induce non-spherical geometry.
If there is a Kerr black hole at its jet base, 
for example following GR-related phenomena may happen;
(i) magneto-spin effect which aligns 
a jet-base along black hole spin, and it leads
to asymmetric geometry (McKinney et al. 2013).
(ii) 
the accretion disk might be warped by 
Bardeen and Peterson effect caused
by the frame dragging effect
(Bardeen and Peterson 1975; Hatchett et al. 1981).
Although a recent research by
Dexter et al. (2012) suggests that
the core emission is not 
dominated by the disk but the jet
component,
the disk emission should be taken into account
if accretion flow emission is largely
blended in the core emission in reality
(see also Broderick and Loeb 2009).
We should take these GR effects
into account when they are indeed 
effective.
(3)
Apart form GR effect,
pure geometrical effect between jet opening angle and 
viewing angle which may cause a partial blending of
SSA thin part of the jet.
It might also cause non-spherical geometry and
inclusion of them is also important.

\end{itemize}


\bigskip
\leftline{\bf \large Acknowledgment}
\medskip

\noindent
We acknowledge the anonymous referee
for his/her careful review and suggestions
for improving the paper.
MK thank A. Tchekhovskoy for useful discussions.
This work is partially supported by 
Grant-in-Aid for Scientific Research,
KAKENHI 24540240 (MK) and 24340042 (AD)
from Japan Society for the Promotion of Science (JSPS).



\begin{figure} 
\includegraphics[width=8cm]{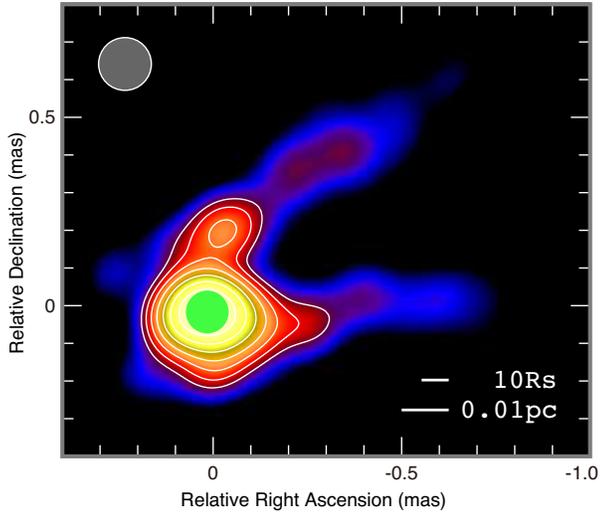}
\caption
{The one-zone sphere region treated
in this work (the yellow-green colored circle)
overlayed on the actual VLBA image of M87 at 43~GHz. 
While the limb-brightening structure is seen 
at the outer part, the central region of 
the radio core can be approximately described as a
uniform sphere.
The diameter of this yellow-green circle corresponds to
$\theta_{\rm FWHM}=0.11~{\rm mas}$. 
Details of the 43~GHz image have been explained in H13.}
\label{fig:m87image_ver2}
\end{figure}%
\begin{figure} 
\includegraphics[width=8cm]{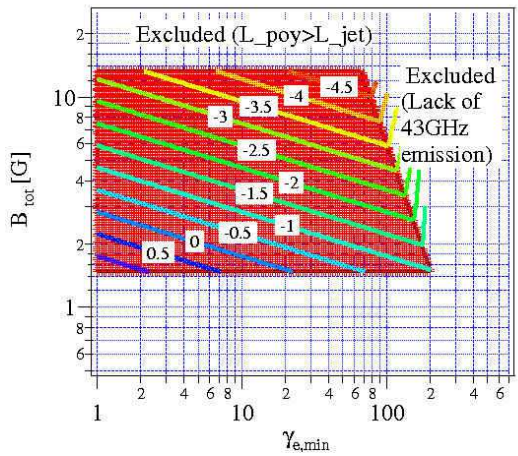}
\caption
{The allowed region in  
$\gamma_{e,\rm min}$ and $B_{\rm tot}$ plane
(the red-colored boxed region)
and the corresponding $\log (U_{e}/U_{B})$ values
with 
$L_{\rm jet} = 5 \times 10^{44}~{\rm erg s^{-1}}$
and $p=3.0$. The $\log (U_{e}/U_{B})$ value is
obtained from Eq.~(\ref{eq:ueub}).
The boundary of the allowed region is 
determined by 
Eqs.~(\ref{eq:gmin}), 
(\ref{eq:energetics}), and 
(\ref{eq:allowed_theta}).
(Short stray lines from the box 
 should be ignored.)}
\label{fig:ueub_5d44_p30}
\end{figure}%
\begin{figure} 
\includegraphics[width=8cm]{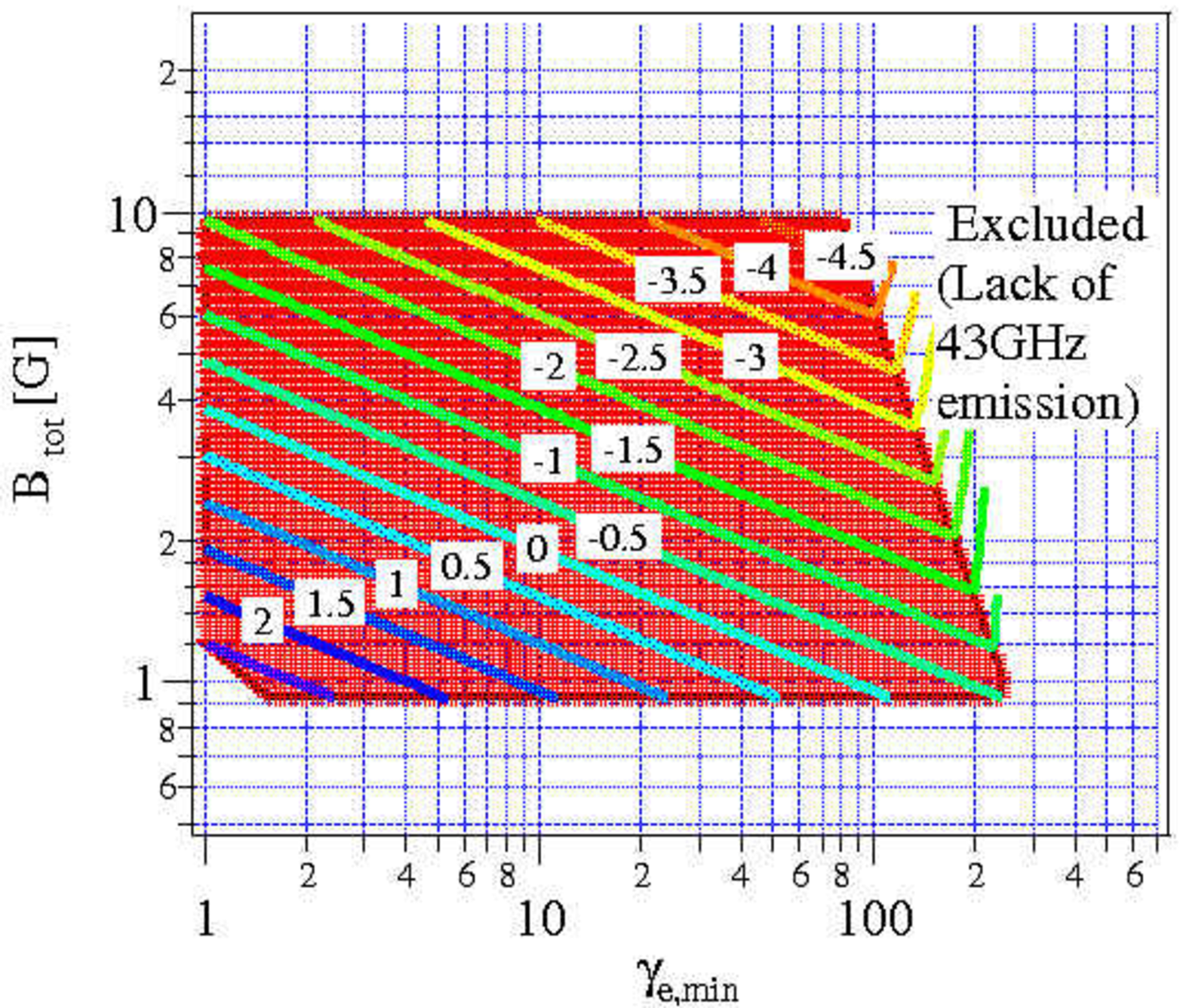}
\caption
{Same as Fig.~(\ref{fig:ueub_5d44_p30}) but with $p=3.5$.}
\label{fig:ueub_5d44_p35}
\end{figure}%
\begin{figure} 
\includegraphics[width=8cm]{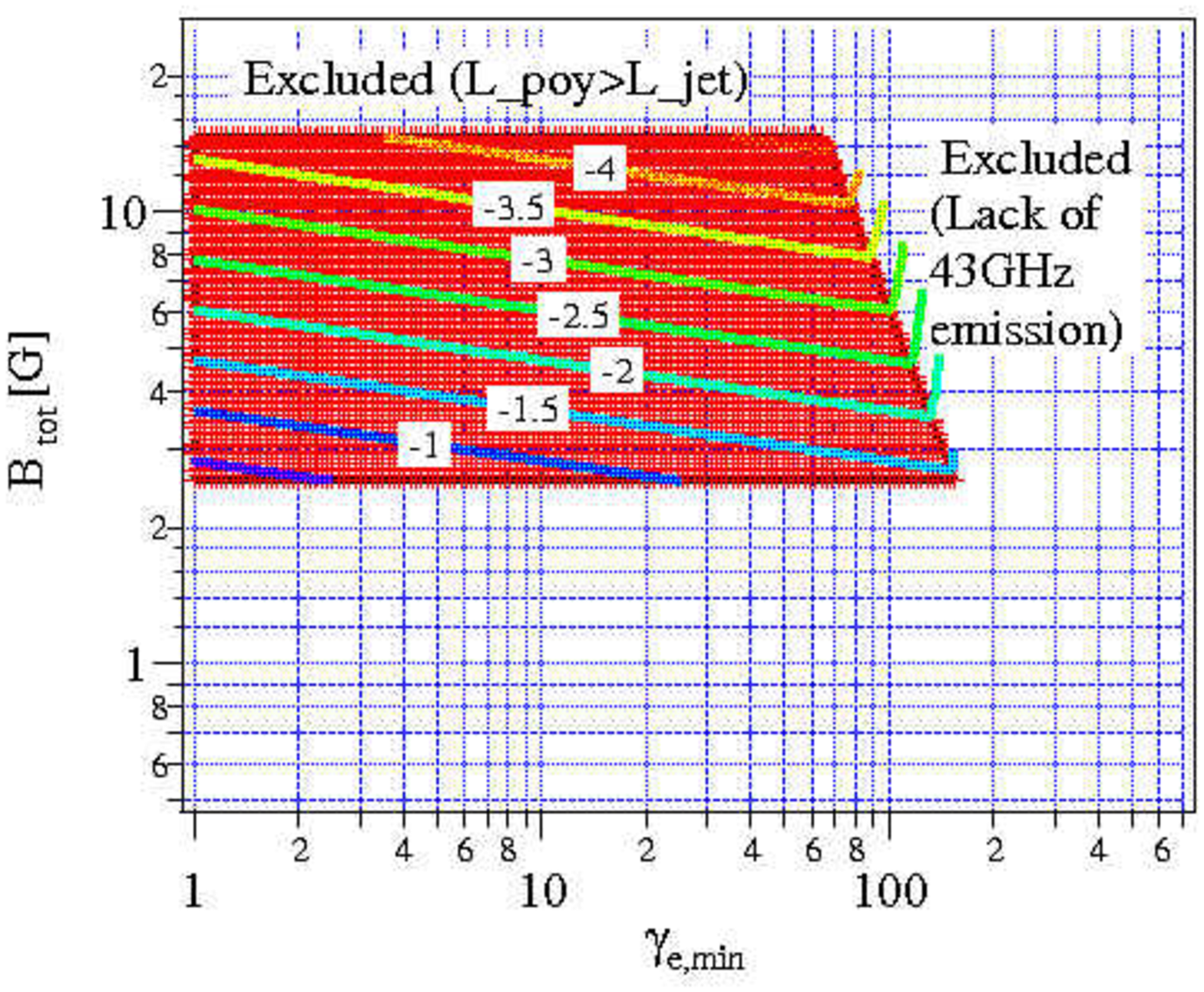}
\caption
{Same as Fig.~(\ref{fig:ueub_5d44_p30}) but with $p=2.5$.}
\label{fig:ueub_5d44_p25}
\end{figure}%

\begin{table}
\centering
\caption{Relevant coefficients for $B$ and $K_{e}$}
\label{table:coefficient}       
\begin{tabular}{lcccccc}
\hline\noalign{\smallskip}
{\bf $p$}    & 
{\bf $c_{1}(p)$} & 
{\bf $c_{2}(p)$} & 
{\bf $b(p)$} & 
{\bf $b(p)$} by Hirotani (2005)&
{\bf $b(p)$} by Marscher (1983)&
{\bf $k(p)$} 
\\
\noalign{\smallskip}\hline\noalign{\smallskip}

 2.5 &
1.516&
0.405&
$3.3\times 10^{-5}$ &
$2.36\times 10^{-5}$ &
$3.6 \times 10^{-5}$ &
$1.4\times 10^{-2}$ 
\\

3.0 &
1.490&
0.303&
$1.9\times 10^{-5}$ &
$2.08\times 10^{-5}$ &
$3.8\times 10^{-5}$ &
$2.3\times 10^{-3}$ 
\\

3.5 &
1.520&
0.245&
$1.2\times 10^{-5}$ &
$1.78\times 10^{-5}$ &
-- &
$3.6\times 10^{-4}$ 
\\

\noalign{\smallskip}\hline
\end{tabular}

\end{table}

\begin{table}
\centering
\caption{Allowed $B_{\rm tot}$ and $\theta_{\rm obs}$}
\label{table:B}       
\begin{tabular}{lcccccc}
\hline\noalign{\smallskip}
{\bf $p$}    & 
{\bf $L_{\rm j}$}    & 
minimum {\bf $B_{\rm tot}$} & 
maximum {\bf $B_{\rm tot}$} &  
minimum {\bf $\theta_{\rm obs}$} & 
maximum {\bf $\theta_{\rm obs}$} & 
\\
  &
[erg~s$^{-1}$] &
[G] &
[G] &
[G] &
[mas]&
\\
\noalign{\smallskip}\hline\noalign{\smallskip}

2.5 &
$1\times 10^{44}$ &
2.5&
7.7&
0.11&
0.15&
\\

2.5 &
$5\times 10^{44}$ &
2.5&
14.7&
0.11&
0.17&
\\

3.0 &
$1\times 10^{44}$ &
1.5&
6.9&
0.11&
0.16&
\\

3.0 &
$5\times 10^{44}$ &
1.5&
13.3&
0.11&
0.19&
\\

3.5 &
$1\times 10^{44}$ &
0.93&
6.3&
0.11&
0.18&
\\

3.5 &
$5\times 10^{44}$ &
0.93&
9.6&
0.11&
0.20&
\\

\noalign{\smallskip}\hline
\end{tabular}
\end{table}

\begin{table}
\centering
\caption{Obtained maximum and minimum $U_{e}/U_{B}$}
\label{table:UeUb}       
\begin{tabular}{lccccc}
\hline\noalign{\smallskip}
{\bf $p$}    & 
{\bf $L_{\rm j}~{\rm [erg~s^{-1}]}$}    & 
{\bf $\max U_{e}/U_{B}$} & 
{\bf $\min U_{e}/U_{B}$} &
\\
\noalign{\smallskip}\hline\noalign{\smallskip}

2.5 &
$1\times 10^{44}$ &
$0.5$&
$3.6\times 10^{-4}$&
\\

2.5 &
$5\times 10^{44}$ &
$0.5$&
$2.4\times 10^{-5}$&
\\

3.0 &
$1\times 10^{44}$ &
$22$&
$1.6\times 10^{-4}$&
\\

3.0 &
$5\times 10^{44}$ &
$22$&
$1.0\times 10^{-5}$ &
\\

3.5 &
$1\times 10^{44}$ &
$1.2\times 10^{2}$&
$0.9\times 10^{-4}$ &
\\

3.5 &
$5\times 10^{44}$ &
$6.1\times 10^{2}$&
$1.3\times 10^{-5}$ &
\\

\noalign{\smallskip}\hline
\end{tabular}
\end{table}

\end{document}